\documentclass[]{interact}
\usepackage{epstopdf}
\usepackage[caption=false]{subfig}
\usepackage{lscape}
\usepackage{multirow}
\usepackage[numbers,sort&compress]{natbib}
\bibpunct[, ]{[}{]}{,}{n}{,}{,}

\theoremstyle{plain}
\newtheorem{theorem}{Theorem}[section]

\theoremstyle{definition}

\theoremstyle{remark}

\begin{document}
\setlength{\textheight}{200mm}
\setlength{\textwidth}{170mm}
\title{Some applications of phase-type distributions in recurrent events}
\author{
\name{Raoufeh Asghari\textsuperscript{a} and Amin Hassan Zadeh\textsuperscript{b} \thanks{CONTACT A. Hassan Zadeh Email: am\_hassanzadeh@sbu.ac.ir} }
\affil{\textsuperscript{a}Department of Statistics, Shahid Beheshti University, Tehran, Iran}
\affil{\textsuperscript{b} Department of Actuarial Science, Shahid Beheshti University, Tehran, Iran}}
\maketitle
\begin{abstract}
In this paper, the recurrent events that can occur more than one over the follow-up time have been modeled by phase-type distributions. We use the finite-state continuous-time Markov process with multi states for patients with recurrent events. The number of recurrences until time $t$, the time stay for every state and the time till death are of importances. The time till death is assumed to have a phase-type distribution (which is defined in a Markov chain environment) with interpretable parameters. The underlying continuous-time Markov chain has one absorbing state (death) and transient states to reflect recovery and disease stages. A system of differential equations is obtained to calculate the probability of various number of transitions, the conditional expected time to stay in a disease stage and the probability of transition from a stage to another. The model has been calibrated via a real and simulated datasets. The bootstrap techniques have been used to construct the confidence intervals for the parameters.
\end{abstract}


\section{Introduction}
Multi-state models are the most common models used for the description of longitudinal survival data. A multi-state model is a model for a stochastic process, which is characterized by a set of states and the possible transitions among them \cite{Andersen1993}. The states represent different situations of the individual (healthy, diseased, etc) along a follow-up. Special multi-state models that have been widely used in biomedical applications are the three-state progressive model, the illness-death model, or the bivariate model \cite{Hougaard2000}. One of the important features of multiple state models is reversible illness-death model that is the model we, in this paper, drive some formulas in a Markovian context.\\
In many biomedical studies, the event of interest can occur more than once over the follow-up time for patients. Such events are called recurrent events \cite{book2012}. The recurrent events are repeated events, which are of the same type such as acute exacerbations in asthmatic children, seizures in epileptics, cancer recurrences, myocardial infarctions, migraine pain, and ear infections. Several statistical models have been proposed in the literature to analyze recurrent events including  \cite{Andersen1982},\cite{Lin2000},\cite{Kelly2000} and \cite{Cook2010}.
\cite{Araujo2014} describes the R package TPmsm which aims at implementing nonparametric and semiparametric estimators for the transition probabilities in progressive illness-death models. Also another R package, TP.idm is introduced in \cite{Balboa2018} that implements a novel non-Markovian estimator for the transition probability matrix in the progressive illness-death model under right censoring. \cite{Meira2009} reviews several modeling approaches following the methodology of the multi-state models, and focusing on the estimation of several quantities such as the transition probabilities and survival probabilities.\\
In this paper, we use a finite-state continuous time Markov process with a single absorbing state. The state space in the Markov process is partitioned to some subspaces. Each subspace will represent a state of a human beings health state. For example, in cancer case, each subspace is interpreted as either a stage of cancer or recovery. Each subspace will include states such that the semi-Markov property can be reflected in the models. See \cite{Asghari2019} for more details.\\

 As in our model the time to death follows a phase-type (PH) distribution, we can take advantages of the PH properties. It is good to remind that the set of \emph{PH} is dense in the class of all distributions defined on the non-negative real numbers, see \cite{Asmussen1996}. There are closed-form expressions for the distribution and density functions, and this also applies to the Laplace--Stieltjes transform and we are able to obtain the expected value and the all non-central moments of the \emph{PH} random variable by successive differentiation of the Laplace transform. \\
Many authors recently have used PH distribution in their researches. \cite{Hassanzadeh2013} uses \textit{PH} for modeling disability insurance. The model represents the aging process as the passage through a number of phases of decreasing vitality. When disabled, individuals additionally pass through several stages that represent duration of disability. \cite{Asghari2019} uses \textit{PH} for  modeling of skin cancer patients in the United States and estimate parameters related to the aging process that can be useful for comparing the physiological age processes of patients with cancer and healthy people. \cite{Lin2007} used phase-type distribution for mortality analysis, and obtained conditional survival probabilities of the time of death and the actuarial present values of the whole life insurance and annuity. \cite{Faddy2000} used phase-type distribution for analyzing data on lengths of stay of hospital patients. \cite{Odd1995} uses \textit{PH} to model the interval time between first and second birth. A special case of \textit{PH} is Coxian distribution that can be used to represent survival times in terms of phases through which an individual may progress until leaves a system, such as hospital stay or time till death. \cite{Marshall2007} used the Coxian to model the patients stay in hospital. The contribution of this paper are presented in Theorem 3.1 and by developing formulas, in the \emph{PH} context, presented in examples. It can be seen that how some formulas in recurrent events context are calculated tractably in a Markov chain, by using the Markov property.\\
 This paper is organized as follows. Section 2 provides an introduction to \textit{PH} as well as the notational convention that will be used in describing our model. In Section 3, we describe the structure of our model and present the main theorem which is the main contribution of this paper. In this section two examples will be presented. Section 4 closes the paper with some concluding remarks and a brief discussion of possible future studies.

\section{Phase-type Distributions}
Consider a continuous-time Markov process $\{J_t,t\ge0\}$ with the space state, $\Gamma=\{0,1,\ldots,m\}$, $m \in \mathbb{N}$, where the state $0$ is absorbing and the rest are transient. Assume the intensity matrix and initial probability vector associated with $J$, are denoted by $\textbf{Q}$ and $\boldsymbol{\beta}$, respectively. Clearly, we have that
  \begin{eqnarray} \label{matrix-Q}
  \textbf{Q}=
  \begin{bmatrix}
    0 & \textbf{0}^\prime \\
    \textbf{t}_0 & \textbf{T}
  \end{bmatrix}
  \end{eqnarray}
where the matrix $\textbf{T}$ is $m\times m$, sub--intensity matrix ( matrix $\mathbf{B}=(b_{ij})_{ i,j=1,\dots,m} $ is called a sub--intensity matrix if  $b_{ii}\leq 0$, $b_{ij} \geq 0, \textrm{and} \, \sum_{k=1}^{m} b_{ik} \leq 0$, with strick inequality for at least on $i$, for ~$i,j=1,2,\dots,m; ~i\neq j$). The vector $\textbf{t}_0$ is the transition rates from the transient states to the absorbing state, 0.  Therefore we have $\textbf{T}\textbf{1}+\textbf{t}_0=\textbf{0}$, (where $\textbf{0}$, $\textbf{1}$  are column vectors of zeros and ones with proper dimensions, respectively). It can be easily proved that
 \begin{eqnarray}
e^{\textbf{Q}y}=
  \begin{bmatrix}
    1 & \textbf{0}^\prime \\
    \textbf{1}-e^{y\textbf{T}}\textbf{1} & e^{y\textbf{T}}
  \end{bmatrix},
\end{eqnarray}
( matrix exponential of the squared matrix $\mathbf{B}$ is defined as $e^{\mathbf{B}}=\sum_{l=0}^{\infty} \frac{\textbf{B}^l}{l !}$).
\newline
We also assume that $P(J_0 \in \{0\}) = \beta_0=0$. As a  result, $\boldsymbol{\beta}= ( \beta_{0}, \boldsymbol{\alpha})$ where we have that $\boldsymbol{\alpha}\textbf{1}=1$.\\
If we define, $Y=\inf\{t; J_t = 0\}$, then it is said that $Y$ is a \emph{PH} random variable with representation $(\boldsymbol{\alpha},\textbf{T})$.
For a continuous random variable $Y$, the event $Y>y$ implies that the process started from any transient state has not reached the absorbing state by time $y$, thus the distribution function of $Y$ can be obtained as follows:\\
\begin{align*}
F(y)=1-\boldsymbol{\alpha}e^{y\textbf{T}}\textbf{1},~~~ y\ge 0.
\end{align*}

Thus the survival function of $Y$ is
\begin{equation} \label{PHSUR}
S(y)=\boldsymbol{\alpha}e^{y\textbf{T}}\textbf{1},~~~y\ge 0.
\end{equation}
 By taking derivatives of $F(y)$, one can obtain the probability density function of $Y$ as follows ,
 \begin{equation*}
f_Y(y)=\boldsymbol{\alpha}e^{y\textbf{T}}\textbf{t}_0,~~~y\ge 0.
\end{equation*}
The Laplace-transform and $k^{th}$ moments are given in the following.\\
$$ \Phi(s)=\boldsymbol{\alpha}(s\textbf{I}-\textbf{T})^{-1}\textbf{t}_0,$$
$$E(Y^k)=k!\boldsymbol{\alpha}(-\textbf{T}^{-1})^k\textbf{1},~~~~k=0,1,2,\cdots.$$
For a complete review of \emph{PH} distributions refer to \cite{bookneuts}.

\section{ Transition probabilities}
Based on the structure of a representation of the PH, several classes of phase-type distributions can be distinguished. The structure of a PH representation often has an impact on its application, as some structures allow more efficient solutions. The most important distinction is the one into Acyclic and General Phase-type distributions: Every acyclic phase-type distribution has at least one Markovian representation without cycles in the sub--generator, while for general phase-type distributions cycles are allowed.
An application of the general  PH is in the reversible illness-death model. In this paper we consider the reversible illness-death model in a Markovian environment.

\begin{figure}
\centering
\includegraphics[scale=0.35]{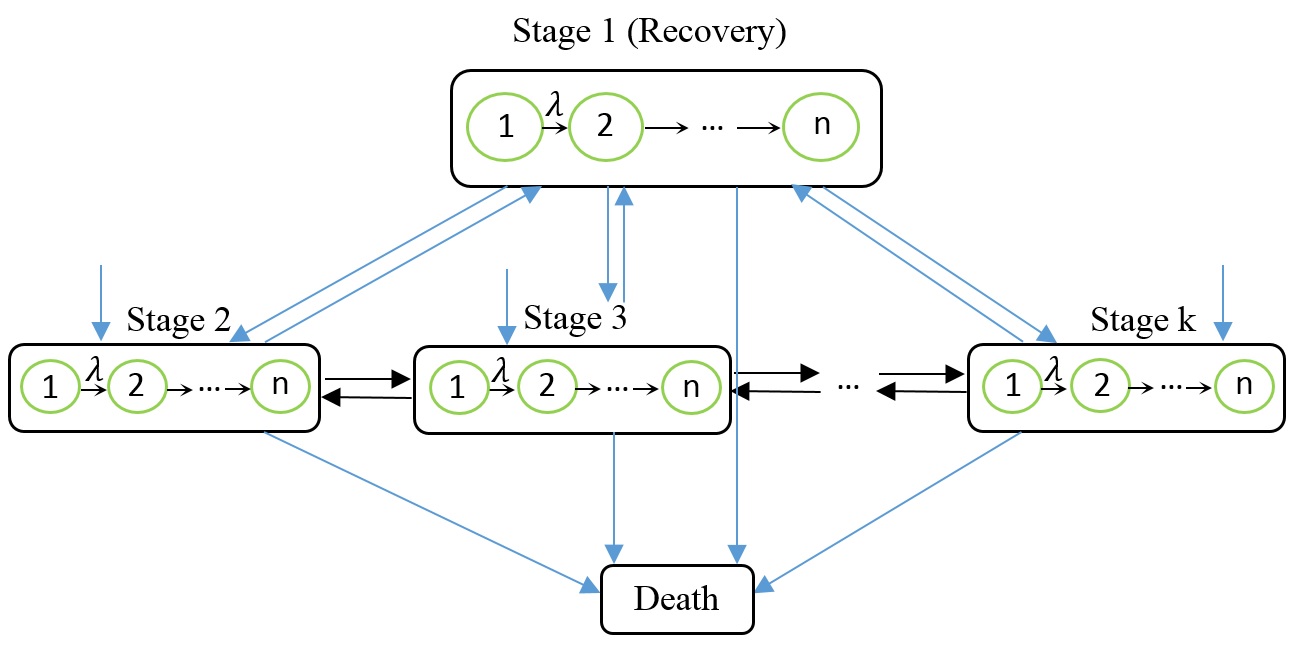}
\caption{Reversible illness-death model: the k+1 stages (boxes) and the possible transition among
them (arrows).} \label{graph1}
\end{figure}

\noindent The analysis in recurrent studies is often performed using the multi--state models. These models are very useful for describing event history data offering a better understanding of the process of the illness, and leading to a better knowledge of the evolution of the disease over time. The complexity of a multi-state model greatly depends on the number of states defined and by the transitions allowed between these states. Our approach is calculating the probability of the number of transitions.\\
We use the finite-state continuous-time Markov process with multi states. One absorbing state which is the death phase and the other states are recovery and disease phases. From now on, we will use the word "stage" rather than "state" or "phase" in our multi-state model. In fact, we will assume $k$ stages ( from 1 to $k$) for the disease and recovery, (one stage for the recovery and $k-1$ for the disease) and one absorbing state for death. Each stage will have $n$ transient states. In this model transition from every stage to another is possible. In some cases the states inside stages can be interpreted as physiological ages, see \cite{Lin2007} and \cite{Asghari2019} where aging is transition from one physiological age to the next physiological age and process will end when transition occurs from any other state to the absorbing state, death. For each state $i, i=1,2,..,n$ in every stage, several parameters are used for modeling mortality. Issues of interest \textit{i.e.}, the number of recurrences or transitions until time $t$, the number of transitions from one stage to another and the expected time stay in every stage will be calculated.\\
After the diagnosis of special disease, that stage is also known. So patient is at one of the disease stages at time $t = 0$ -arrival time- at age $x$. At any time thereafter, he may arrive at death state, recovery or visit another disease stages. Recovered patient may become sick or dead. Our proposed multi-state model is reversible, in the sense that past stages can be revisited. For this model, let $E$ denote the space state of the underlying Markov chain, then $E=\bigcup_{j=1}^k E_i \bigcup D$ where sets of $E_1,E_2,\dots.E_k$ are to represent stages, including recovery, and the state $D=\{0\}$ represents death. Every stage, including recovery, has a finite set of $n$ states labeled $1, ..., n$, with instantaneous transitions being possible between selected pairs of states, see Figure \ref{graph1}.
In general, there are $k\times n + 1$ states, where $k \times n$ are for recovery and disease stages and 1 for the death state.\\
For each $t \geq 0$, the continuous Markovian random process $J_t$ is in one of the stages $ 1, ..., k$ or in $D$. We interpret the event $J_t \in E_i$ to mean that the individual is in stage $i$ at age $x + t$, $i =1,\dots,k$.
We assume the rate of transition from the disease stages depend on stage, age and duration of illness. Therefore, we design our Markovian model such that it reflects the semi-Markovian property.\\
As mentioned earlier, we propose a reversible illness--death model involving $n$ states in each alive stage. By doing so, the semi-Markov property can be captured in the model. This method have been used in  \cite{Hassanzadeh2013}. Patients may move to the next stage of disease, recover or death. 
The recovery rate will normally be lower for the later stages of disease. Of course, patients may die while in any state of stages, as indicated by the arrows to death in Figure \ref{graph1}.\\
Since the Markov process has only a single absorbing state, the time of death (the time of absorption) follows a \textit{PH} distribution. Note also that the time staying in recovery or disease statuses can be obtained by using the Markov property.\\
Furthermore, the sub-intensity matrix, $\textbf{T}$, of the Markov process in \eqref{matrix-Q} is given by
\begin{equation}\label{T}
\textbf{T}=
  \begin{bmatrix}
    \textbf{T}_1 & \textbf{T}_{1,2} & \textbf{T}_{1,3} & \cdots & \textbf{T}_{1,k} \\
    \textbf{T}_{2,1} & \textbf{T}_2 & \textbf{T}_{2,3} & \cdots & \textbf{T}_{2,k} \\
    \vdots & \vdots & \vdots & \vdots & \vdots \\
    \textbf{T}_{k,1} & \textbf{T}_{k,2} & \textbf{T}_{k,3} & \cdots & \textbf{T}_{k}
  \end{bmatrix},
\end{equation}
where $\textbf{T}_{i,j}, i \neq j,~ i,j=1,2,...,k$ are $n \times n$ matrices containing the transition rates from the stage $i$ to $j$. The matrices $\textbf{T}_i$'s, $i =1,2,...,k$ are the sub-intensity matrices (with non-zero main and upper diagonal elements) for the Markov chain describing a sojourn in stage $i$. The mortality rates, $\mathbf{t}_0=-\textbf{T}\boldsymbol{1}$, is an $n\times k$-dimensional column vector containing the rate of death from each of the $n\times k$ alive states.\\
In this paper, we assume that the person at diagnose has age of $x$ and we will remove $x$ from the notations.\\
From now on, it as also assumed that
$\boldsymbol{\alpha}=(\boldsymbol{\alpha_1},\boldsymbol{\alpha_2},\dots,\boldsymbol{\alpha_{k}})$, where $\boldsymbol{\alpha_i}$'s are all $1\times n$ vectors corresponding to initial probabilities to start from stage $i$ ($\boldsymbol{\alpha_1}$ corresponds to the recovery which can be assumed to be $\boldsymbol{0}$). The normalized and the transposed vectors of a vector $\boldsymbol{\beta}$ are denoted by $\hat{\boldsymbol{\beta}}$ and $\boldsymbol{\beta}'$, respectively. \\
Now with the notations above and assumptions, by using the Markov property we are able to calculate the probability of various number of transitions until time $t$.\\
\begin{theorem}\label{THM1}: Let $\{J_t,t\geq 0\}$ be a continuous-time Markov chain with space state $E=\bigcup_{j=1}^{k}E_j\cup D$ and with the sub-intensity matrix \eqref{T} where $D$ the only absorbing state and $E_j$'s are mutual disjoint subsets of $E$. Assume $\{N(t),t\geq 0\}$ is the number of transitions between $E_j$'s in $[0,t]$ or from any $E_j$ to $D$, $j=1,...,k$, and define $P^{(i)}_{t}(l)=P[N(t)=l|J_0\in E_i]$. Then we have the followings:\\
\begin{equation}\label{PN0}
P^{(i)}_t(0)=\hat{\boldsymbol{\alpha}}_i\,e^{t\textbf{T}_i}\textbf{1}
\end{equation}
\begin{equation}\label{PN1}
P^{(i)}_t(1)=\sum_{\substack{i_1 =1 \\ i_1\neq i }}^{k}\hat{\boldsymbol{\alpha}}_i\,\mathbf{x}^{(i)}_{i_1}(t)\textbf{1},
\end{equation}
\begin{equation}\label{PN2}
P^{(i)}_t(2)=\sum_{\substack{i_1,i_2\\ i_1\neq i\\i_2\neq i_1 }}^{k}\hat{\boldsymbol{\alpha}}_i \mathbf{x}^{(i)}_{i_1,i_2}(t)\textbf{1},
\end{equation}
$$\vdots$$
\begin{equation}\label{PNk}
P^{(i)}_t(l)=\sum_{\substack{i_1,i_2,...,i_l\\ i_1\neq i\\i_2\neq i_1\\ \vdots\\i_l\neq i_{l-1}}}^{l}\hat{\boldsymbol{\alpha}}_i \mathbf{x}^{(i)}_{i_1,i_2\cdots, i_l}(t)\textbf{1}
\end{equation}
where $\mathbf{x}^{(i)}_{i_1}(.),\mathbf{x}^{(i)}_{i_1i_2}(.),\cdots, \mathbf{x}^{(i)}_{i_1i_2\cdots i_k}(.)$ satisfy the following differential equations:
$$\frac{d}{dt}\mathbf{x}^{(i)}_{i_1}(t)=e^{\textbf{T}_i\,t}\textbf{T}_{i,i_1}+\mathbf{x}^{(i)}_{i_1}(t)\textbf{T}_{i_1},$$

$$\frac{d}{dt}\mathbf{x}^{(i)}_{i_1,i_2}(t)=\mathbf{x}^{(i)}_{i_1}(t)\textbf{T}_{i_1,i_2}+\mathbf{x}^{(i)}_{i_1,i_2}(t)\textbf{T}_{i_2},$$ and
$$\frac{d}{dt}\mathbf{x}^{(i)}_{i_1,i_2,\cdots,i_l}(t)=\mathbf{x}^{(i)}_{i_1,i_2\cdots i_{l-1}}(t)\textbf{T}_{i_{l-1},i_l}+\mathbf{x}^{(i)}_{i_1,i_2,\cdots,i_l}(t)\textbf{T}_{i_l},$$
with the matrices $\textbf{x}$ satisfy the following recursive equations
$$\textbf{x}_{i_1,i_2,\dots,i_{m-1},i_m}^{(i)}(t)=\int_{0}^t \textbf{x}_{i_1,i_2,\dots,i_{m-1}}^{(i)}(z)\textbf{T}_{i_{m-1}i_m}e^{\textbf{T}_{i_m}(t-z)}dz,$$
for $m=1,2,\dots,$ where $$\textbf{x}_{\{\}}^{(i)}(t)=e^{\textbf{T}_i\,t},$$
and all $\textbf{x}$'s are zero matrices, with proper dimension, at $t=0$.\\
\end{theorem}
\noindent \textbf{Proof}: See Appendix 1.\\\\
To solve the differential equations above, we use ODE-45 function in MATLAB (2018) software. In the following we will consider two examples for the applications.\\

\begin{figure}
	\centering
	\includegraphics[scale=0.5]{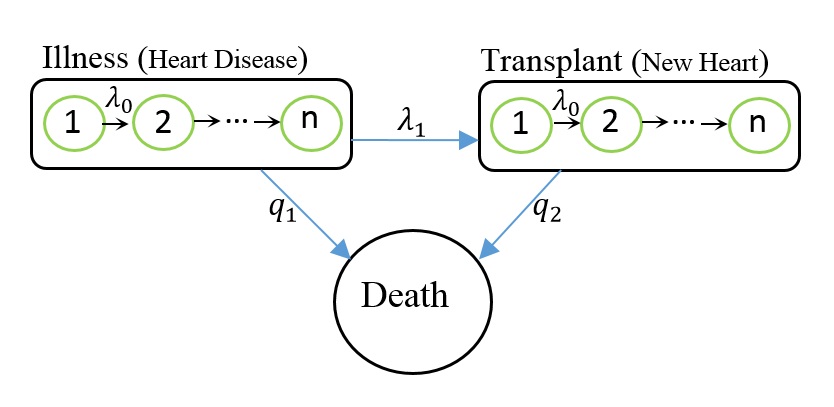}
	\caption{Illness-death model with disease and transplant states.} \label{graph4}
\end{figure}

\noindent \textbf{Example~1}: The Stanford Heart Transplantation Study\\
The Stanford heart transplant study began in October $1967$. This data set can be found in \cite{Kalb1980} (pp 230-232) or in \cite{Crowley1977}. The available data covers the period until April 1, $1974$. Some patients died before an appropriate heart was found. Out of the $103$, $69$ received a heart transplant that 45 (65\%)of them deceased. Of the 34 patients without transplant, the number of deaths was 30 (88\%). The remaining $28$ alive patients contributed with censored survival times. For each individual, an indicator of its final vital status (censored or not), the survival times (time to transplant, time to death) from the entry of the patient into the study (in days), and a vector of covariates including age at acceptance (Age), year of acceptance (Year), previous surgery (Surgery: coded as $1$ = yes; $0$ = no), and transplant (Transplant: coded as $1$ = yes; $0$ = no) were recorded. The Transplant covariate is the only time-dependent covariate.\\
In this data-structure, an individual's survival data is expressed by three variables: start, stop and event. For the Stanford study, the time-dependent covariate “transplant” represents a treatment intervention. Individuals without change in the time-dependent covariate are represented by only one line of data, whereas patients with a change in the time-dependent covariate must be represented by two lines. For these patients, the first line represents the time period until the transplant; the second line represents the time period that passes from the transplant to the end of the follow-up or death. The remaining (time-fixed) covariates are the same for the two lines. For each row, variables start and stop mark the time interval (start, stop) for the data, while event is an indicator variable taking on value $1$ if there was a death at time stop, and 0 otherwise. As an example consider the information available from four patients (from the Stanford study) with identification numbers $25$, $26$, $27$ and $28$ in Table \ref{tab4} . For the first two patients the time from enrollment to censoring is $1800$ and $1401$ days, respectively, and the first patient had a heart transplant $25$ days after enrollment. The time from enrollment to death for the third and fourth patients are $263$ and $72$ days, respectively, and the last patient received a new heart in day $71$.\\
\begin{table}
	\caption{ Stanford heart transplantation in Example 2.}
	\label{tab4}
	\begin{tabular}{lccccccc}
		\hline
		id &start &stop&event&transplant&age&year&surgery \\
		\hline
		25 &0 &25 &0 &0 &33.2238 &1.57426 &0\\
		25 &25 &1800 &0 &1 &33.2238 &1.57426 &0\\
		26 &0 &1401 &0 &0 &30.5353 &1.58248 &0\\
		27 &0 &263 &1 &0 &8.7858 &1.59069 &0\\
		28 &0 &71 &0 &0 &54.0233 &1.68378 &0\\
		28 &71 &72 &1 &1 &54.0233 &1.68378 &0\\
		\hline
	\end{tabular}
\end{table}
The descriptive statistics of the data by Age (intervals of 5-year), Year (from 1967 to 1974, intervals of 1) and Surgery (yes=1 or no=0) are presented in Tables \ref{tab4.1}, \ref{tab4.2} and \ref{tab4.3}, respectively.
\begin{table}
	\caption{ The descriptive statistics of the data by age in Example 2.}
	\label{tab4.1}
	\begin{tabular}{l|cccccc}
		\hline
		Age & N & Trans. & Death &$\textrm{Death}_{\textrm{trans.}}$&\% $\textrm{Death}_{\textrm{without trans.}}$&\% $\textrm{Death}_{\textrm{after trans.}}$ \\
		\hline
		$<$25 &5 &2 &4 &1&100&50\\
		25-30 &5 &4 &1 &1&0&25\\
		30-35 &4 &2 &1 &0&50&0\\
		35-40 &7 &4 &4 &1&100&25\\
		40-45 &19&12&15&9&100&75\\
		45-50 &29&22&19&13&86&59\\
		50-55 &27&17&23&14&90&82\\
		$>$55 &8 &6 &8 &6&100&100\\
		\hline
		sum &103 &69 &75 &45\\
	\end{tabular}
\end{table}
In Table \ref{tab4.1}, the average age of heart patients is 45.2 years. The percent of transplanted patients with 45-50 years old is maximum with 78\% transplanting. The rate of death for transplanted patients($\frac{\# \text{death of transplanted patients}}{\#\text{ transplanted patients}}$) increases with age.\\
\begin{table}
	\caption{ The descriptive statistics of the data by year in Example 2.}
	\label{tab4.2}
	\begin{tabular}{l|cccccc}
		\hline
		Year & N & Trans. & Death &$\textrm{Death}_{trans.}$&\% $\textrm{Death}_{\textrm{without trans.}}$& \% $\textrm{Death}_{\textrm{after trans.}}$\\
		\hline
		1 &16 &7 &16 &7&100&100\\
		2 &17 &11 &15 &10&83&91\\
		3 &10 &8 &6 &4&100&50\\
		4 &17 &11 &14 &8&100&73\\
		5 &17 &12 &12 &8&100&67\\
		6 &19 &16 &11 &8&100&50\\
		$>$6 &7 &4 &1 &0&33&0\\
		\hline
		sum &103 &69 &75 &45\\
	\end{tabular}
\end{table}
In Table \ref{tab4.2}, the number of heart patients' enrollment is about the same for every year but the percent of transplanted patients for the fifth year of study is maximum with 84\% transplanting. The mortality rate for transplanted patients has decreased over the years so it is expected that the influence of year of acceptance on hazard is negative.\\
\begin{table}
	\caption{ The descriptive statistics of the data by surgery in Example 2.}
	\label{tab4.3}
	\begin{tabular}{l|cccccc}
		\hline
		surgery & N & Trans. & Death &$\textrm{Death}_{\textrm{trans.}}$&\% $\textrm{Death}_{\textrm{without trans.}}$&\% $\textrm{Death}_{\textrm{after trans.}}$\\
		\hline
		0 &87 &56 &66 &39&87&70\\
		1 &16 &13 &9 &6&100&46\\
		\hline
		sum &103 &69 &75 &45\\
	\end{tabular}
\end{table}
\noindent As can be seen in Table~\ref{tab4.3}, most patients have no history of surgery and patients without surgery are more likely to die, about 76\%. Heart patients with surgery are more likely to have heart transplants than those without heart surgery, in other words, 81\% of patients with surgery and 64\% of patients without surgery have a heart transplant. Therefore it is expected that the influence of previous surgery on hazard is negative.\\
In this example, we use the illness-death model that can be used to study the effect of binary time-dependent covariates, shown in Figure \ref{graph4}.
The heart patients are in the illness status when the disease is diagnosed. At any time thereafter, they may become transplanted or deceased. As explained in Section 1, rates of transition from the illness state depend on the duration of disease. Therefore the model must be a semi-Markov model.\\
We propose a model involving $2$ stages (Illness and Transplant) as shown in Figure \ref{graph4}. In each stage, the patients may move from left to right in the figure as they age. After a heart transplant, patients move right from a state in the illness stage to a state in the transplant stage. Of course, patients may die while in any state, as indicated by the arrows to death state in Figure \ref{graph4}.\\
In order to describe how probabilities values are calculated, we define some notations for our model. For $t \geq 0$, let $J_t$ represent the status of a patient at time $t$. We use the finite-state continuous-time Markov process with $2n$ alive states: $n$ transient states for disease stage, $n$ transient states for transplant stage, and one absorbing state which is the death state. Suppose that the space state is partitioned into the set $E_1$ of $n$ illness states and the set $E_2$ of $n$ transplant states and $D$, for death.\\
Hence the sub-intensity matrix \eqref{T} will be reduced to \begin{equation}\label{matrix_T}
\mathbf{T}=
\begin{bmatrix}
\textbf{T}_1 & \textbf{T}_{12} \\
\textbf{0} & \textbf{T}_2
\end{bmatrix},
\end{equation}
where the $\textbf{T}_1$ and $\textbf{T}_2$ are the sub-intensity (with non-zero main and upper diagonal elements) matrices for the Markov chain describing a sojourn in $E_1$, and $E_2$, respectively. The matrix $\mathbf{T}_{12}$ contains the transition rates from the illness stage to the transplant stage. Let $\boldsymbol{\alpha}$ be an 2n-dimensional column vector providing the initial state probabilities. \\

The elements of the proposed sub-intensity matrix $\textbf{T}$ are:

\begin{equation}
\textbf{T}=
\begin{bmatrix}
-(q^1_1+\lambda_0+\lambda_1)& \lambda_0 &\cdots& 0 & \lambda_1 & 0 & 0&\cdots \\
0 & -(q^1_2+\lambda_0+\lambda_1) & \lambda_0 &\cdots& 0 & \lambda_1 & 0&\cdots \\
\vdots & \vdots & \ddots &\vdots& \vdots & \vdots& \ddots & \vdots \\
0 & 0 &\cdots& -(q^1_n+\lambda_1)& 0 &0 &\cdots & \lambda_1\\
0 & 0 &\cdots& 0 & -(q^2_1+\lambda_0) & \lambda_0 & 0&\cdots\\
0 & 0 &\cdots& 0 & 0 & -(q^2_2+\lambda_0) & \lambda_0&\cdots\\
\vdots & \vdots & \ddots &\vdots& \vdots & \vdots& \ddots & \vdots \\
0 & 0 &\cdots& 0 & 0 & 0 &\cdots& -q^2_n\\
\end{bmatrix},
\end{equation}
where $\lambda_0$ is the rate of transition from one state to the next, $\lambda_1$ is the rate of transplant from disease.
The rate of mortality for the stages decease and transplant are shown by $\mathbf{q}^1_i$ and $\mathbf{q}^2_i$, respectively.
\begin{eqnarray} \label{qi}
\mathbf{q}^1_i&=&a+b+q\times i^{(p+\gamma_1age+\gamma_2year+\gamma_3surgery)} \nonumber \\
\mathbf{q}^2_i&=&a  +q\times i^{(p+\gamma_1age+\gamma_2year+\gamma_3surgery)}
\end{eqnarray}
for $i=1,2,\dots,n$. Where the constant $a$ is interpreted as a background rate that a general reflection of the living environment, $q$ is a scale parameter and $p$ is a measure of the relative impact of the occurrence of aging. By descriptive statistics from heart data, rate of death for heart patients, before transplant is higher than after. So we add the parameter $b$ to the elements of $\mathbf{q}^1_i$. The three regression coefficients, $\gamma_1, \gamma_2$, and $\gamma_3$ are included in the model to represent Age, Year, Surgery effects.\\
Also it was assumed that patients at acceptance will be in the first phase of the model. That is, $\boldsymbol{\alpha}=(1,0,..,0)^\prime$.\\

We aim to estimate the parameters by the maximum likelihood method. In the heart transplant data, there are four situations for a patient: He stays in disease stage (stage 1) until the end of the study period, dies before the transplant, dies after the transplant or he is alive after transplant (stage 2) until end of the follow-up. Since the Markov process has only a single absorbing state, the time of death (the time of absorbing) follows a PH distribution. In order to estimate the parameters of the model, by using the phase-type density function \eqref{PHSUR} with representation ($\boldsymbol{\alpha},\mathbf{T}$), we can construct expressions for the contribution of the likelihood function.\\

1. The probability of staying in disease state for a period of time $t$:
\begin{equation}\label{prob_1}
f_1=Pr[J_{u}\in E_1~~\forall u\in [0,t]|J_0 \in E_1]=\boldsymbol{\hat{\alpha}}_1^\prime e^{t\textbf{T}_1}\textbf{1}.
\end{equation}

2. The probability of staying in disease state for $s$ time units and then stay in transplant state until time $t$, will be:
\begin{equation}\label{prob_2}
f_{12}= Pr[\underset{\forall u\in [0,s]}{J_{u}\in E_1},\underset{\forall v\in [0,t-s]}{J_{v+s}\in E_2}|J_0 \in E_1]= \boldsymbol{\hat{\alpha}}_1^\prime \, e^{\textbf{T}_1\,s}\textbf{T}_{12} e^{\textbf{T}_2 (t-s)}\textbf{1}.
\end{equation}

3.  The probability of death in disease state after $t$ time units is:
\begin{equation}\label{prob_3}
f_{10}=Pr[\underset{\forall u\in [0,t)}{J_{u}\in E_1},J_{t}\in D|J_0 \in E_1 ]=\boldsymbol{\hat{\alpha}}_1^\prime e^{\textbf{T}_1\,t}\textbf{t}_{10}.
\end{equation}
where $\textbf{t}_{10}=-\textbf{T}_1 \textbf{1}$.\\

4.  The probability of death at time $t$ from the transplant after staying in disease state for $s$ time units is:
\begin{equation}\label{prob_4}
\begin{split}
f_{20}&=Pr[ \underset{\forall u\in [0,s]}{J_{u}\in E_1}\,\underset{\forall \nu \in [0,t-s]}{J_{\nu+s}\in E_2},J_{t}\in D|J_0 \in E_1]\\&=\boldsymbol{\alpha}_1^\prime\,e^{s\textbf{T}_1}\textbf{T}_{12}e^{\textbf{T}_2(t-s)}\textbf{t}_{20}
\end{split}
\end{equation}

where $\mathbf{t}_{20}=-\textbf{T}_{2} \textbf{1}$.
We estimate the parameters in \eqref{qi} by maximizing the likelihood function, $$L(\theta)=f_{1}^{n_1}f_{12}^{n_2}f_{10}^{n_3}f_{20}^{n_{4}},$$
where $n_k$s, $k=1,2,3,4,$ represent the number of patients in each scenarios explained in \eqref{prob_1}-\eqref{prob_4} and    $\theta=(a,~b,~q,~p,~\lambda_0,~\lambda_1,~\gamma_1,~\gamma_2,~\gamma_3)$.\\
The FMINCON function in MATLAB Software has been used for finding the maximum likelihood estimates.
We estimated the parameters for different numbers of $n$. Based on the results shown in Table \ref{tab4.5}, $n=3$ was chosen. Note that the our criterion is maximizing $l(\theta)=log(L(\theta))$. \\
\begin{table}
	\caption{Parameters values for different $n$ values.}
	\label{tab4.5}
	\begin{tabular}{lcccccccccc}
		\hline
		$n$& $a$ & $q$ & $p$&$\lambda_0$&$\lambda_1$& $b$&$\gamma_1$&$\gamma_2$&$\gamma_3$&$l(\theta)$\\
		\hline
		1 & \small{1.7e-03}&\small{5.0e-06} & \small{7.5}& \small{1.50}& \small{0.0116}& \small{0.0033}& \small{1.0} &\small{-0.25} &\small{-0.54 }& \small{-896.48}\\
		2 & \small{1.7e-03}&\small{6.7e-08}& \small{3.0}& \small{0.50}& \small{0.0116}& \small{0.0034}& \small{0.37} &\small{-0.17} &\small{-0.59} &\small{ -896.45}\\
		3 & \small{4.9e-04}& \small{6.4e-08} & \small{9.3}&\small{0.50}& \small{0.0115}& \small{0.0034}& \small{0.098} &\small{-0.02} &\small{-0.92} & \small{-885.17}\\
		4 & \small{7.7e-04}& \small{2.7e-08} & \small{7.7}& \small{0.49}& \small{0.0113}& \small{0.0034}& \small{0.104} &\small{-0.003} &\small{-0.98} &\small{ -885.88}\\
		5 & \small{7.7e-04}& \small{1.2e-08} & \small{7.1} &\small{0.49}& \small{0.0115}& \small{0.0036}& \small{0.094} &\small{-1.2e-07} &\small{-0.98} & \small{-886.25}\\
		6 & \small{8.4e-04}&\small{8.7e-09} & \small{6.5} & \small{0.49}& \small{0.0116}& \small{0.0037}& \small{0.089} &\small{-0.015} &\small{-0.87} & \small{-886.78}\\
		8 & \small{1.0e-03}& \small{5.9e-10} & \small{6.7}& \small{0.30}& \small{0.0116}& \small{0.0037}& \small{0.094 }&\small{-2.1e-08} &\small{-0.86 }&\small{ -888.37}\\
		10& \small{8.9e-04}& \small{5.1e-09 }& \small{5.2} & \small{0.48}&\small{0.0119}& \small{0.0038}& \small{0.077 }&\small{-1.3e-08} &\small{-0.75} & \small{-887.63}\\
		20& \small{1.7e-03}& \small{7.5e-09 }& \small{2.2} &\small{0.008}&\small{0.0116}& \small{0.0034}& \small{0.49 }&\small{-0.47} &\small{-0.60} & \small{-895.97}\\
		\hline
	\end{tabular}
\end{table}
The parameters values, the standard deviations and the confidence intervals at level 95\% are shown in Table \ref{tab5}.\\
In the fitted model, the influence of Age at acceptance on hazard is positive, while effects of Year and Surgery are both negative as they were expected.\\
The standard deviation of each parameter is obtained by the bootstrap technique. The main benefit of the bootstrap is that it allows statisticians to set the confidence intervals of the parameters without having to make unreasonable assumptions. It creates multiple resamples (with replacement) from a single set of observations, and computes the effect size of interest on each of these resamples. The bootstrap resamples of the effect size can then be used to determine the $95\%$ confidence interval. See \cite{EfTib1993} for more about bootstrap techniques.\\
\begin{table}
	\small
	\caption{Parameters values with standard deviations and confidence intervals($95\%$).}
	\label{tab5}
	\begin{tabular}{lcccccccccc}
		\hline
		Param.& a & q & p&$\lambda_0$&$\lambda_1$& b&$\gamma_1$&$\gamma_2$&$\gamma_3$&$sup_\theta l(\theta)$\\
		\hline
		Estimate&\small{4.9e-04}& \small{6.4e-08} & \small{9.3}&\small{0.50}& \small{0.0115}& \small{0.0034}& \small{0.098} &\small{-0.02} &\small{-0.92} & \small{-885.17}\\
		Std. &              \small{5.3e-04}&\small{2.9e-06} &\small{2.36}&\small{0.34} &\small{ 0.0036}&\small{0.0018}&\small{0.49}&\small{0.37} &\small{0.22}& \\
		\small{(Lower,~~)}&\small{1.9e-04}&\small{9.2e-14}&\small{1.5e-08}&\small{4.8e-04}&\small{0.007}&\small{4.9e-04}&\small{0.055}&\small{-0.50}&\small{-0.98}&\\
		\small{(~~,Upper)} &\small{0.0025} &\small{1.0e-05}&\small{9.7}   &\small{1.25} &\small{0.021} &\small{0.0077} &\small{1.77}&\small{-1.8e-08}&\small{-0.0014}&\\
		\hline
	\end{tabular}
\end{table}
It was our interest to know about the significance of the covariates and the parameter $b$ in the model. To this end, we have conducted hypothesis testing for some null tests. In Table \ref{tab6} the results are presented.\\
\begin{table}
	\caption{The value of estimates}
	\label{tab6}
	\begin{tabular}{lcccccccccc}
		\hline
		Parameter & a & q & p&$\lambda_0$&$\lambda_1$& b&$\gamma_1$&$\gamma_2$&$\gamma_3$&$sup_\theta l(\theta)$\\
		\hline
		\small{without $\gamma_1$,$\gamma_2$,$\gamma_3$}& \small{0.0017}&\small{4.3e-09}&\small{3.01}&\small{0.50}&\small{0.0116}&\small{0.0033}& -& -& - & \small{-896.48} \\
		\small{without $\gamma_1$}&\small{0.0013}&\small{9.9e-06}&\small{5.1}&\small{0.50}&\small{0.0114}&\small{0.0029}&-&\small{-0.50}&\small{-0.96}&\small{-895.54}\\
		\small{without $\gamma_2$}&\small{6.1e-04}&\small{1.2e-08}&\small{10.4}&\small{0.21}&\small{0.0114}&\small{0.0036}&\small{0.14}&-&\small{-0.98}&\small{-885.02}\\
		\small{without $\gamma_3$}&\small{5.0e-04}&\small{3.5e-07}&\small{7.4}&\small{0.50}&\small{0.0116}&\small{0.0036}&\small{0.13}&\small{-3.7e-04}&-&\small{-888.01}\\
		\small{without b}&\small{0.0022} &\small{2.1e-09} &\small{0.39} & \small{0.0035}& \small{0.0112 }& -&\small{1.71}&\small{-0.077} &\small{ -0.52} & \small{-904.7}\\
		\hline
	\end{tabular}
\end{table}
\newline
The likelihood ratio tests ($LRT$) has been used for the testing. For example, for the null hypothesis  $H_0:\forall i,\gamma_i=0$ against the alternative $H_1: \exists i,\gamma_i\neq 0 $ can be expressed:
\begin{equation}
LRT=-2log(\frac{sup_{\Theta_0} L(\theta)}{sup_{\Theta} L(\theta)})=-2(sup_{\Theta_0} l(\theta)-sup_{\Theta} l(\theta))=22.62,
\end{equation}
which is greater than $ \chi^2_{0.95,3}=7.8$ and therefore the null hypothesis is disproved. As seen in Table \ref{tab6}, we can obtain the values of $LRT$ for the model without every variable separately. $\Theta$ is the parameter space and $\Theta_0$ is the parameter under the null hypothesis. \\
For testing the null hypothesis $H_0:b=0$ against the alternative $H_1: b\neq 0$, $LRT$ is $39.06$, by comparing with $\chi^2_{0.95,1}=3.8$, we reject $H_0$. Hence, it can be said that mortality before heart transplant is greater than after transplant.\\

Now using the value of parameters, the probability of the number of transition until time $t$ can be calculated. These probabilities are shown in Table \ref{tab7} for different Age ( $30$ and $50$), Year (years $3$ and $5$) and Surgery ($0$ and $1$). As we can seen in this table, the probability without transition decreases as $t$ increases because the patient may die or transplant after more time of staying in disease stage. The event of no transition decreases with age, increases with year and surgery. \\
As we said before, mortality increases with age, so probability of staying in the disease stage for older patients are reduced, and he/she is more likely to die. Therefore, it is expected that the probability of one transition increases with age. Mortality decreases with year and surgery, so the probability for no transition also increases as the year and surgery increase. The probability of one transition increases until 6 months but decreases thereafter because the probability of transplant decreases in the later months and the probability of death increases. The probability of two transitions constantly increases. It is obvious that this probability increases with age because of its positive influence on death, decreases with year and surgery because of their negative influence on death.\\
\begin{table}
	\caption{Probability of N(t) for different Age, Year and Surgery.}
	\label{tab7}
	\begin{tabular}{l|cccccccc}
		\hline
		Age&Year&Surgery &P[N(t)] &1 month & 3 months & 6 months & 1 year & 3 years \\
		\hline
		\multirow{12}{*}{30}&\multirow{6}{*}{3} &\multirow{3}{*}{0}&P[N(t)=0] & 0.6254 & 0.2441 & 0.0595 & 0.0033  & 3.5e-08\\
		&&&P[N(t)=1] & 0.3714 & 0.7338 & 0.8786 & 0.8505  & 0.6082\\
		& &&P[N(t)=2] & 0.0032 & 0.0221 & 0.0619 & 0.1462  & 0.3918\\\cline{4-9}
		&&\multirow{3}{*}{1}&P[N(t)=0] & 0.6279 & 0.2474 & 0.0612 & 0.0035  & 4.1e-08\\
		& &&P[N(t)=1] & 0.3695 & 0.7350 & 0.8892 & 0.8776  & 0.6650\\
		& &&P[N(t)=2] & 0.0026 & 0.0176 & 0.0496 & 0.1190  & 0.3350\\\cline{2-9}
		
		& \multirow{6}{*}{5}&\multirow{3}{*}{0}&P[N(t)=0] & 0.6255 & 0.2443 &  0.0596 & 0.0033 & 3.5e-08\\
		&&&P[N(t)=1] & 0.3713 & 0.7339 & 0.8793 & 0.8523 & 0.6117\\
		&&&P[N(t)=2] & 0.0032 & 0.0218 & 0.0611 &  0.1444 & 0.3883\\\cline{4-9}
		& &\multirow{3}{*}{1}&P[N(t)=0] & 0.6280 & 0.2475 & 0.0612 & 0.0035 & 4.2e-08\\
		&&&P[N(t)=1] & 0.3695 & 0.7350 & 0.8894 & 0.8783 & 0.6666\\
		&&&P[N(t)=2] & 0.0026 & 0.0175 & 0.0493 & 0.1183 & 0.3334\\
		\hline
		\multirow{12}{*}{50}&\multirow{6}{*}{3} &\multirow{3}{*}{0}&P[N(t)=0] & 0.5955 & 0.2077 & 0.0428 & 0.0017 & 4.5e-09\\
		&&&P[N(t)=1] & 0.3935 & 0.7215 & 0.7764 & 0.6360 & 0.3831\\
		&&&P[N(t)=2] & 0.0110 & 0.0709 & 0.1808 & 0.3624 & 0.6169\\\cline{4-9}
	 &&\multirow{3}{*}{1}&P[N(t)=0] & 0.6168 & 0.2332 & 0.0542 & 0.0027 & 2.0e-08\\
		 & &&P[N(t)=1] & 0.3777 & 0.7299 & 0.8453 & 0.7712 & 0.4802\\
		& &&P[N(t)=2] & 0.0055 & 0.0369 & 0.1005 & 0.2261 & 0.5198\\\cline{2-9}
		
		& \multirow{6}{*}{5}&\multirow{3}{*}{0}&P[N(t)=0] & 0.5969 & 0.2093 & 0.0434 & 0.0017 & 5.0e-09\\
	&&&P[N(t)=1] & 0.3925 & 0.7220 & 0.7804 & 0.6426 & 0.3844\\
		&&&P[N(t)=2] & 0.0106 & 0.0688 & 0.1762 & 0.3557 & 0.6156\\\cline{4-9}
		& &\multirow{3}{*}{1}&P[N(t)=0] & 0.6173 & 0.2339 & 0.0545 & 0.0027 & 2.0e-08\\
		&&&P[N(t)=1] & 0.3774 & 0.7301 & 0.8472 & 0.7756 & 0.4859\\
		&&&P[N(t)=2] & 0.0053 & 0.0360 & 0.0982 & 0.2217 & 0.5141\\
		\hline
	\end{tabular}
\end{table}
Given $J_u \in E_i$, the time of staying continuously in stage $i$ of disease has a PH distribution with representation $(\boldsymbol{\hat{\alpha}}^u_i, \textbf{T}_i)$, where
$$\boldsymbol{\hat{\alpha}}^u_i=\dfrac{\boldsymbol{\hat{\alpha}}^\prime\,e^{\textbf{T}u}\textbf{I}_{E\,E_i}}{\boldsymbol{\hat{\alpha}}^\prime\,e^{\textbf{T}u}\textbf{1}_i},$$
$\textbf{I}_{E\,E_c}$, $c=1,\dots,k$, is $nk \times n$ matrix with $(n(c-1)+l,l)$ entry equal to 1, $l=1,\dots,n$ and all other entries are 0, and $\textbf{1}_c$ is a column vector with $nk$ entries such that elements $n(c-1)+l$, $l=1,\dots, n$ equal to $1$ and all other entries $0$.

 \noindent Hence, given that the process $J$ is $i^{\textrm{th}}$ stage at time $u$, the expected time of continuously staying in this stage, in time interval $[0,t]$ is equal to:

\begin{equation}\label{EN0}
\begin{split}
E[\int_{0}^{t}\mathbf{1}_{\{J_\nu \in E_i,\,\forall \nu \in [0,\omega] | J_u\in E_i\}}d\omega]&=\int_{0}^{t}Pr[J_\nu\in E_i,\,\forall \nu \in [0,\omega] | J_u\in E_i]\, d\omega\\
&=\int_{0}^{t} \hat{\boldsymbol{\alpha}}^{u}_i\,e^{\textbf{T}_i\,\omega}\textbf{1}_i\, d\omega\\
&=\boldsymbol{\hat{\alpha}}^{u}_i\,\textbf{T}_i^{-1}(e^{\textbf{T}_i\,t}-\textbf{I})\textbf{1}.
\end{split}
\end{equation}

The integrand in \eqref{EN0} can be interpreted as the survival function of a PH distribution with representation of $(\boldsymbol{\hat{\alpha}}^u_i, \textbf{T}_i)$.
Given starting from stage illness (stage 1), expected times without transition for $u=0$, are shown in Table \ref{tab8} for different ages.
It is expected that the 30-year-old patient spends 24 days in disease stage during one month. This expected time decrease as age increases because of transition to death state.\\
\begin{table}
	\caption{Expected sojourn time (days) of without transition}
	\label{tab8}
	\begin{tabular}{l|ccccc}
		\hline
		    &  \multicolumn{5}{c}{duration}  \\
		Age & 1 month & 3 months & 6 months & 1 year & 3 years\\
		\hline
		30 & 24 & 48.3 &  60.1 & 63.6 & 63.9\\
		40 & 23.9 & 47.6 &  58.7 & 61.9 & 62.1\\
		50 & 23.5 & 45.6 &  55.0 & 57.3 & 57.4\\
		60 & 22.5 & 40.5&  46.3 & 47.3 & 47.3\\
		\hline
	\end{tabular}
\end{table}
Finally, the probability of transition from stage $i$ to stage $j$ in a period of time $t$ can be calculated easily as follows
\begin{equation}\label{transfromi2j}
Pr[J_{u+t} \in E_j| J_u \in E_i]=\dfrac{\hat{\boldsymbol{\alpha}}^\prime e^{\textbf{T}u}\textbf{I}_{E\,E_i}\textbf{I}^\prime_{E\,E_i}e^{\textbf{T}t}\textbf{1}_j }{\hat{\boldsymbol{\alpha}}^\prime e^{\textbf{T}u}\textbf{1}_i}.
\end{equation}

For Age$=30$, Year$=3$ and Surgery$=0$, and $u=0$ \eqref{transfromi2j} can be seen in Table \ref{tab9}. The formula for the probability of transition from disease stage to transplant is greater than death. Mortality from disease and transplant stages increase over time but mortality before transplant is greater than after.
\begin{table}
	\caption{Probability of transition from stage $i$ to stage $j$ for Age=30, Year=3, Surgery=0.}
	\label{tab9}\label{i2j}
	\begin{tabular}{lcccccc}
		\hline
		from &  to & 1 months & 3 months & 6 months & 1 year & 3 years \\
		\hline
		disease& transplant & 0.2726 & 0.5338 & 0.6296 & 0.5866 & 0.3434\\
		disease& death      & 0.0988 & 0.2000 & 0.2490 & 0.2640 & 0.2648\\
		transplant& death   & 0.0032 & 0.0221 & 0.0619 & 0.1462 & 0.3918\\
		\hline
	\end{tabular}
\end{table}
\newpage
\noindent \textbf{Example~2}: Cancer Disease\\
After someone is diagnosed with cancer, doctors will try to figure out if it has spread, and if so, how far. This process is called staging. The stage of a cancer describes how much cancer is in the body. It helps determine how serious the cancer is and how best to treat it. Cancer staging may sometimes include the grading of the cancer. This describes how similar a cancer cell is to a normal cell.
Doctors also use a cancer's stage when talking about survival statistics. The earliest stage of cancers are called stage 0 (carcinoma in situ), and then range from stages I (1) through IV (4). As a rule, the lower the number, the less the cancer has spread. A higher number, such as stage IV, means cancer has spread more. Although each person's cancer experience is unique, cancers with similar stages tend to have a similar outlook and are often treated in much the same way. When cancer returns after a period of remission, it's considered a recurrence. A cancer recurrence happens because, in spite of the best efforts to rid of cancer, some cells from cancer remain. These cells could be in the same place where cancer first originated, or they could be in another part of body. These cancer cells may have been dormant for a period of time, but eventually they continued to multiply, resulting in the reappearance of the cancer.\\
Also by the end of the observation period, the patient under study may not have reached an absorbing state. In survival analysis this would correspond to individual still being alive by the end of the study and this kind of incomplete observation is known as right-censoring.\\
To illustrate the computation of probabilities of transitions, we use  hypothetical transition rate between stages. In the following, we show how the PH model can be used in modeling of recurrent events in cancers.\\ 
The model can be considered as a development of \cite{HassanZadehetalldis}, where the number of disease stages (which in \cite{HassanZadehetalldis} is called disability stage) is increased from 2 to 5. \\  
We assume $5$ disease stages and one recovery stage. Each disease stage and recovery include 5 states, \textit{i.e.} $k=6$ and $n=5$. The stage $1$ corresponds to the recovery (R) and the stages $2,3,\dots,6$ represents cancer stages $0,1,\dots,4$, respectively. The entries of the sub-generator matrix $\textbf{T}=(t_{ij}) \,\, i,j=1,...,30$ will be described in the following. \\ 

The rates of transition from one state to the next are given by
$$t_{i,i+1}=\lambda, \,\,\, i=1,2,...,29, i\neq 5,10,15,20,25,$$ 
where $\lambda$ is assumed to be $0.2$.\\
The rates of recovery from disease stage, defined based on the degree of malignancy of the disease, are
$$t_{i+5l,i-5}=\gamma(0.1)^l, ~~~ i=6,7,...,10, ~l=0,1,...,4,$$
where $\gamma$ is the coefficient of recovery; assume $\gamma=0.1$\\
and the rates of transition from one stage of disease to the stage before, defined based on the degree of malignancy of the disease, are given by
$$t_{i+5l,i+5(l-1)}=\beta(0.1)^{l},~~~ i=11,12,...,15, ~l=0,...,3$$
where $\beta$ is assumed to be $0.2$.\\
The rates of transition from one stage of disease to the next are
$$t_{i+5l,i+5(l+1)}=(l+1)(a+qi^p),~~~i=1,2,...,5,~l=1,...,4$$
with $a=10^{-3}$, $q=10^{-6}$ and $p=4.5$.\\
The rates of transition from recovery stage to each of the disease stages are given by
$$t_{i,i+5(l+1)}=(0.1)^l(a+qi^p),~~~i=1,2,...,5,~l=0,1,...,4,$$
whereas the mortality rates are given by
$$\mathbf{t}^\prime_0=(t_{1,0},t_{2,0},...,t_{30,0})$$
where
$$t_{i,0}=(0.1)^5+a+qi^p,~~~i=1,...,5$$
and for $l=0,1,...,4$
$$t_{i+5(l+1),0}=(0.1)^{4-l}+a+qi^p,~~~i=1,...,5.$$
So other entries of $\textbf{T}$ are zero except the diagonal entries which are:
$$t_{i,i}=-\sum_{j=0,j\neq i}^{30}t_{i,j}, ~~~ i=1,2,...,30.$$
Finally, the initial probability is $\boldsymbol{\alpha}'=(\boldsymbol{\alpha}^\prime_1,\boldsymbol{\alpha}^\prime_2,...,\boldsymbol{\alpha}^\prime_6)$ where $\boldsymbol{\alpha}_i$'s are $5$-dimensional column vector. After the diagnosis of cancer, first stage, $i^{\textrm{th}}$, is also known so $\boldsymbol{\alpha}^\prime_i$ is $(1,0,0,0,0)$ and other elements of $\boldsymbol{\alpha}$ will be zero.\\
\begin{table}
	\caption{Probability of N(t) in Example 2.}
	\label{table1}
	\begin{tabular}{lccccc}
		\hline
		 &  \multicolumn{5}{c}{duration}  \\
		& input stage & 6 months & 12 months & 24 months & 36 months \\
		\hline
		P(N(t)=0) &  &  &  &  & \\
		&0 &0.5382&0.2885&0.0814&0.0226\\
		&1 &0.2750&0.0752&0.0055&0.0004\\
		&2 &0.8046&0.6429&0.3981&0.2403\\
		&3 &0.5219&0.2701&0.0698&0.0175\\
		&4 &0.0025&6.04e-6&3.62e-11&2.16e-16\\
		\hline
		P(N(t)=1) &  &  &  &  & \\
		&0 & 0.4541 &0.6880& 0.8504 & 0.8576\\
		&1 & 0.5199 &0.4422 & 0.1960 & 0.0952\\
		&2 & 0.1406 &0.2061 & 0.2694 & 0.2959\\
		&3 & 0.4573 &0.6915 &0.8703  & 0.9132\\
		&4 & 0.9975 & 0.9999& 0.9998 & 0.9998\\
		\hline
		P(N(t)=2) &  &  &  &  & \\
		&0 & 0.0088 &0.0204 & 0.0440 &0.0669 \\
		&1 & 0.2875 &0.4614 & 0.3268 & 0.1523\\
		&2 & 0.0639 &0.1173 & 0.1483 & 0.1552\\
		&3 & 0.0208 &0.0377 & 0.0559 & 0.0622\\
		&4 & 7.94e-5 & 1.31e-4& 1.71e-4 & 1.81e-4\\
		\hline
	\end{tabular}
\end{table}
\noindent Table \ref{table1} shows the probability of $N(t)$, \eqref{PN0}--\eqref{PN2}, for different input stages, for different durations. 
These probabilities behave the way we expect. With entering to every stage, probability of no transition is decreasing as the duration increases. 
This means that probability of staying in every stage is higher at the earlier times and decreases over time. 
For example in stage $4$, probability is near to zero over time, because stay in this state very low and patient has a higher chance to die. 
For one transition until $t$ this probability is near to one, because of the transition to death state.
As it is seen in Table \ref{table1}, the probability of staying in stage $2$ is the maximum, compared to other stages, due to the fact 
that chance of recovery is not as high as in stage 1 and the rate of death is not as high as in stage 3.\\
The expected time of staying in stage at diagnosis, \eqref{EN0}, is shown in Table \ref{table2}. When a patient is diagnosed with stage $0$ cancer,
 it is expected that he stays in that stage for 4.5 months during the six months. If the first stage is 2, this value will be 5.4 months. As we said before, at this stage, the expected sojourn time is greater
than the others because of the low chance of recovery compared to the previous stage and low chance of death compared to the next stage. The patient, who is in stage 4, is expected to eventually survive for 1 month.\\
\begin{table}
	\caption{Expected sojourn time in first stage until $t$.}
	\label{table2}
	\begin{tabular}{lcccc}
		\hline
		& \multicolumn{4}{c}{duration}\\
	input stage & 6 months & 12 months & 24 months & 36 months \\
		\hline
		0 &4.5 &6.9 &8.8 &9.4\\
		1 &3.4 &4.3 &4.6 &4.6\\
		2 &5.4 &9.7 &15.9 &19.6\\
		3 &4.4 &6.7 &8.5 &9.0\\
		4 &1 &1 &1 &1\\
		\hline
	\end{tabular}
\end{table}
Also we can obtain the probability of one transition from stage $i$ to stage $j$, $P[N_{ij}(t)=1]$, by using Equation \eqref{transfromi2j}.
\begin{table}
	\caption{Probability of one transition between stages.}
	\label{tab3}
	\begin{tabular}{lccccccccc}
		\hline
		stage &waiting time &$R$&$0$&$1$&$2$&$3$&$4$&$D$ \\
		\hline
		0 & 6 months& 0.4440 & 0      & 0.0050 & 0      & 0      & 0 & 0.0051 \\
		&12 months& 0.6750 & 0      & 0.0046 & 0      & 0      & 0 & 0.0084 \\
		1 & 6 months& 0.0334 & 0.4704 & 0      & 0.0092 & 0      & 0 & 0.0069 \\
		&12 months& 0.0420 & 0.3809 & 0      & 0.0104 & 0      & 0 & 0.0089 \\
		2 & 6 months& 0.0054 & 0      & 0.0592 & 0      & 0.0165 & 0 & 0.0596 \\
		&12 months& 0.0096 & 0      & 0.0635 & 0      & 0.0247 & 0 & 0.1084 \\
		3 & 6 months& 4.38e-04 & 0   & 0      & 0.0078 & 0      & 0.0032 & 0.4459 \\
		&12 months& 6.58e-04 & 0   & 0      & 0.0103 & 0      & 0.0021 & 0.6784 \\
		4 & 6 months& 9.89e-6 & 0   & 0      & 0      & 1.16e-4& 0    & 0.9973 \\
		&12 months& 9.72e-6 & 0   & 0      & 0      & 6.049e-5& 0    & 0.9998 \\
		\hline
	\end{tabular}
\end{table}
As seen in Table \ref{tab3}, probability of one transition from the lower stages of cancer to recovery stage is higher than the higher stages, this means that the lower the number, the less the cancer has spread and the higher number of cancer has spread more. In higher stages probability of transition to recovery is near to zero. Also probability of transition from the lower stages of cancer to death is lower than the higher stages and probability of transition from higher stages to death is near to one. Other possible probabilities from one stage to another is shown in this table.\\\\

\section{Conclusion}
In this article, the multi-state models, that are widely used in the medical studies for many diseases, has been developed via taking advantages of phase--type distribution, to model the recurrent events. It is important for decision makers and patients to be aware of information about their illness, such as the probability of relapse, the time stay in each stage of recovery or disease, the probability of recovery, and so on. Using Markov's properties and phase-type distributions, we present a formula for calculating the probability of the number of transitions between the stage os diseases. To calculate these probabilities, which are interdependent and defined as recursive relations, we need to use differential equations to calculate them. In this article, we used two examples to calibrate our model. Example $1$ was for 103 Stanford heart patients who have two stages: heart disease and heart transplant. The mortality rate was defined for these stages, which have several parameters. We estimate these parameters by the maximum likelihood method to estimate the matrix \textbf{T}. The standard deviations and the confidence intervals for the parameters are obtained by the bootstrap technique. Finally, the probability of the number of transition until time $t$, the expected time of continuously staying in each stage and the probability of transition from stage $i$ to stage $j$ in a period of time $t$ are obtained. Another example is a simulated cancer that has one recovery stage and five cancer stages. In the simulated example, which had a more complex model, similar calculations have been done.\\

\noindent Calculating these probabilities for higher recurrent's number is time consuming due to more transition between stages. Because of using of the recursive equations in the formulas, we need to use the differential equations of the previous equations to calculate the probability. In other words, we have to start from the beginning to get the probability. \\
We have used the available functions in Matlab and have not developed any algorithms for solving the differential equations. In future study, we aim to optimize the algorithms for calculating probability distribution of the transitions.

\newpage
\section*{Data Availability} 
Data availability statement: NA

\newpage

\section{Appendix~1}

In this section we prove Theorem \ref{THM1}. The Equation \eqref{PN0} is obvious. To prove the rest,
we divide $[0,t]$ into $m$ equal subintervals, each length will be $ds= \frac{t}{m}$.
First, we assume that any transition between the stages will occur in a multiplier of $ds$. Assume also that $s_\kappa=\kappa\,ds$, $\kappa=0,1,\dots$, then we have the followings:
\begin{equation*}
\begin{split}
P^{(i)}_t(1)&=Pr[N(t)=1|J_0\in E_i]=E[\mathbf{1}_{N(t)=1}|J_0\in E_i]\\
&=E[\mathbf{1}_{\{\bigcup_{\substack{i_1 =1 \\ i_1\neq i }}^{k}\bigcup_{\kappa=0}^{m-1} {\underset{\forall u\in [0,s_\kappa]}{J_u\in E_i},~ J_{s_\kappa+ds}\in E_{i_1},~ \underset{\forall \nu \in [0,t-(s_\kappa+ds)]}{J_{s_\kappa+ds+\nu}\in E_{i_1}} }\}}|J_0\in E_i]\\
&=E[\sum_{\substack{i_1 =1 \\ i_1\neq i }}^{k}\sum_{\kappa=0}^{m-1}\,\mathbf{1}_{\{\underset{\forall u\in [0,s_\kappa]}{J_{u}\in E_i} , J_{s_\kappa+ds}\in E_{i_1},\underset{\forall \nu \in [0,t-(s_\kappa+ds)]}{J_{s_\kappa+ds+\nu}\in E_{i_1}}\}}|J_0\in E_i]\\
&=\sum_{\substack{i_1 =1 \\ i_1\neq i }}^{k}\sum_{\kappa=0}^{m-1}\,
E[\mathbf{1}_{\{\underset{\forall u\in [0,s_\kappa]}{J_{u}\in E_i} , J_{s_\kappa+ds}\in E_{i_1},\underset{\forall \nu \in [0,t-(s_\kappa+ds)]}{J_{s_\kappa+ds+\nu}\in E_{i_1}} \}}|J_0\in E_i]\\
&=\sum_{\substack{i_1 =1 \\ i_1\neq i }}^{k} \sum_{\kappa=0}^{m-1} Pr[\underset{\forall u\in [0,s_\kappa]}{J_{u}\in E_i},J_{s_\kappa+ds}\in E_{i_1},~ \underset{\forall \nu \in [0,t-(s_\kappa+ds)]}{J_{s_\kappa+ds+\nu}\in E_{i_1}}|J_0\,\in E_i]\\
&=\sum_{\substack{i_1 =1 \\ i_1\neq i }}^{k} \sum_{\kappa=0}^{m-1} Pr[\underset{\forall u\in [0,s_\kappa]}{J_{u}\in E_i}| J_0\in E_i]Pr[J_{s_\kappa+ds}\in E_{i_1} | J_{s_\kappa}\in E_i] Pr[\underset{\forall \nu \in [0,t-(s_\kappa+ds)]}{J_{s_\kappa+ds+\nu}\in E_{i_1}}|J_{s_\kappa+ds}\in E_{i_1}].
\end{split}
\end{equation*}
For $i,i_1=1,...,k,D$, $i\neq i_1$.\\
Now, by using equation \eqref{PN0}, and letting $ds \to 0$ we end up with the followings:
\begin{equation}\label{appen_proof1}
\begin{split}
=\sum_{\substack{i_1 =1 \\ i_1\neq i }}^{k}\lim_{ds\rightarrow 0}\sum_{\kappa=0}^{m-1}(\hat{\boldsymbol{\alpha}}_i e^{\textbf{T}_i\kappa\,ds}\textbf{I}_{E,E_i}^\prime)(e^{\textbf{T}ds}\textbf{I}_{E,E_{i_1}})(e^{\textbf{T}_{i_1}(t-(\kappa+1)ds)}\textbf{1})\\
\end{split}
\end{equation}
from $\lim_{\epsilon\rightarrow 0^+}\frac{e^{\textbf{T}\epsilon}-\textbf{I}}{\epsilon}=\textbf{T}$, the continuity of $e^{\textbf{T}s}$ and the limit of the Riemann sum, \eqref{appen_proof1} becomes
\begin{equation}\label{appen_proof_2}
\begin{split}
&=\sum_{\substack{i_1 =1 \\ i_1\neq i }}^{k}\int_{0}^{t}\hat{\boldsymbol{\alpha}}_i e^{\textbf{T}_i\,s}\textbf{T}_{i,{i_1}}e^{\textbf{T}_{i_1}(t-s)}\textbf{1}ds\\
&=\sum_{\substack{i_1 =1 \\ i_1\neq i }}^{k}\hat{\boldsymbol{\alpha}}_i \int_{0}^{t}e^{\mathbf{T}_i\,s}\textbf{T}_{i,i_1}e^{\textbf{T}_{i_1}(t-s)}ds\textbf{1}\\
&=\sum_{\substack{i_1 =1 \\ i_1\neq i }}^{k}\hat{\boldsymbol{\alpha}}_i \textbf{x}^{(i)}_{{i_1}}(t)\textbf{1}.
\end{split}
\end{equation}
Where $\textbf{x}^{(i)}_{{i_1}}(t)=\int_{0}^{t}e^{\textbf{T}_i\,s}\textbf{T}_{i,i_1}e^{\textbf{T}_{i_1}(t-s)}ds$. It can be seen that $$\frac{d}{dt}\textbf{x}^{(i)}_{{i_1}}(t)=e^{\textbf{T}_i\,t}\textbf{T}_{i,i_1}+\textbf{x}^{(i)}_{{i_1}}(t)\textbf{T}_{i_1}.$$
The event of $N(t)=2$ means there is $2$ transitions during interval $[0,t]$. Assuming the transitions happen at time $s_{\kappa_1}$ and $s_{\kappa_2}$, we have the following:

\begin{equation*}
\begin{split}
P^{(i)}_t(2)&=Pr\left(N(t)=2|J_0\in E_i\right)=E\left(\mathbf{1}_{N(t)=2}|J_0\in E_i\right)\\
&=E\left(\mathbf{1}_{\bigcup_{\substack{i_1,i_2\\ i_1 \neq i_2 }}^k\bigcup_{\kappa_1=0}^{m-2}\bigcup_{\kappa_2=\kappa_1+1}^{m-1}\underset{\forall~u\in [0,s_{\kappa_1}]}{\,J_{u}\in E_i}\,~J_{s_{\kappa_1}+ds}\in E_{i_1},~\underset{~\forall~\nu\in[0,s_{\kappa_2}-(s_{\kappa_1}+ds)]}{J_{s_{\kappa_1}+ds+\nu}\in E_{i_1}},~J_{s_{\kappa_2}+ds}\in E_{i_2},\underset{~\forall~\omega \in [0,t-(s_{\kappa_2}+ds)]}{J_{s_{\kappa_2}+ds+\omega}\in E_{i_2}}}\right)\\
&=\sum_{\substack{i_1,i_2\\ i_1\neq i\\i_2\neq i_1 }}^{k} \sum_{\kappa_1=0}^{m-2}\sum_{\kappa_2=\kappa_1+1}^{m-1}Pr[\underset{\forall u\in [0,s_{\kappa_1}]}{J_{u}\in E_i}|J_0 \in E_i]\,Pr[J_{s_{\kappa_1}+ds}\in E_{i_1} | J_{s_{\kappa_1}}\in E_i] \\
& Pr[\underset{\forall \nu\in [0,s_{\kappa_2}-(s_{\kappa_1}+ds)]}{J_{s_{\kappa_1}+ds+\nu}\in E_{i_1}}| J_{s_{\kappa_1}+ds}\in E_{i_1}]\,Pr[J_{s_{\kappa_2}+ds}\in E_{i_2}|J_{s_{\kappa_2}}\in E_{i_1}]\\
&Pr[\underset{\forall~\omega \in [0,t-(s_{\kappa_2}+ds)]}{J_{s_{\kappa_2}+ds+\omega}\in E_{i_2}}|J_{s_{\kappa_2}+ds}\in E_{i_2}].
\end{split}
\end{equation*}	

As $ds \to 0$ and using the same techniques used in \eqref{appen_proof_2} we will end with the following.

\begin{equation*}
\begin{split}
&=\sum_{\substack{i_1,i_2\\ i_1\neq i\\i_2\neq i_1 }}^{k}\int_{0}^{t}\int_{0}^{z}\hat{\boldsymbol{\alpha}}_i e^{\textbf{T}_i\,s}\textbf{T}_{i\,i_1}e^{\textbf{T}_{i_1}(z-s)}\textbf{T}_{{i_1}{i_2}}e^{\textbf{T}_{i_2}(t-z)}\textbf{1}~dsdz\\
&=\sum_{\substack{i_1,i_2\\ i_1\neq i\\i_2\neq i_1 }}^{k}\hat{\boldsymbol{\alpha}}_i\int_{0}^{t}\int_{0}^{z} e^{\textbf{T}_i\,s}\textbf{T}_{i\,i_1}e^{\textbf{T}_{i_1}(z-s)}\textbf{T}_{{i_1}{i_2}}e^{\textbf{T}_{i_2}(t-z)}\textbf{1}~dsdz\\
&=\sum_{\substack{i_1,i_2\\ i_1\neq i\\i_2\neq i_1 }}^{k}\hat{\boldsymbol{\alpha}}_i \int_{0}^{t} \mathbf{x}_{i_1}^{(i)}(z)\textbf{T}_{{i_1}{i_2}}e^{\textbf{T}_{i_2}(t-z)}\textbf{1}dz\\
&=\sum_{\substack{i_1,i_2\\ i_1\neq i\\i_2\neq i_1 }}^{k}\hat{\boldsymbol{\alpha}}_i \mathbf{x}_{i_1\,i_2}^{(i)}(t)\textbf{1}.
\end{split}
\end{equation*}
Where $\mathbf{x}_{i_1\,i_2}^{(i)}(t)$ satisfies the following differential equation
$$\dfrac{d\mathbf{x}_{i_1\,i_2}^{(i)}(t)}{dt}=
\mathbf{x}_{i_1}^{(i)}(t)\textbf{T}_{i_1\,i_2}+\mathbf{x}_{i_1\,i_2}^{(i)}(t)\textbf{T}_{i_2}.$$
In general, it can be seen easily that $ P^{(i)}_t(l)$ equals
\begin{equation*}
\begin{split}
&=\sum_{\substack{i_1,i_2,\dots,i_l\\ i_1\neq i\\i_2\neq i_1 \\ \vdots \\ i_{l} \neq i_{l-1}}}^{k}\int_{0}^{t}\int_{0}^{z_{l-1}}\dots\,\int_{0}^{z_{1}}\hat{\boldsymbol{\alpha}}_i e^{\textbf{T}_i\,s}\textbf{T}_{i\,i_1}e^{\textbf{T}_{i_1}(z_1-s)}\textbf{T}_{{i_1}{i_2}}
e^{\textbf{T}_{i_2}(z_2-z_1)} \dots \textbf{T}_{i_{l-1}i_{l}} e^{\textbf{T}_{i_l}(t-z_{l-1})} \textbf{1}~ds\,dz_1\,dz_2\,\cdots dz_{l-1}\\
&=\sum_{\substack{i_1,i_2\\ i_1\neq i\\i_2\neq i_1 }}^{k}\hat{\boldsymbol{\alpha}}_i \int_{0}^{t}
\mathbf{x}_{i_1,i_2,\cdots,i_{l-1}}^{(i)}(z_{l-1})\textbf{T}_{{i_{l-1}}{i_l}}e^{\textbf{T}_{i_l}(t-z_{l-1})}\textbf{1}dz_{l-1}\\
&=\sum_{\substack{i_1,i_2\\ i_1\neq i\\i_2\neq i_1 }}^{k}\hat{\boldsymbol{\alpha}}_i
\mathbf{x}_{i_1,i_2,\cdots,i_l}^{(i)}(t)\textbf{1},
\end{split}
\end{equation*}
where $\mathbf{x}_{i_1,i_2,\cdots,i_l}^{(i)}(t)$ satisfies the following differential equation
$$\dfrac{d\mathbf{x}^{(i)}_{i_1,i_2,\cdots,i_l}(t)}{dt}=\mathbf{x}^{(i)}_{i_1,i_2\cdots i_{l-1}}(t)\textbf{T}_{i_{l-1},i_l}+\mathbf{x}^{(i)}_{i_1,i_2,\cdots,i_l}(t)\textbf{T}_{i_l}.$$

\begin{thebibliography}{9}
	
	\bibitem{Hassanzadeh2013}
	\textsc{Hassanzadeh, A.}, \textsc{Jones, B. L.} and \textsc{Stanford, D. A.} (2013). \textit{The use of phase-type models for disability insurance calculations}, Scandinavian Actuarial Journal \textbf{2014}: 714--728.
	
	\bibitem{book2012}
	\textsc{Kleinbaum, D. G.} and \textsc{Klein, M.} (2012). \textit{Survival Analysis. A Self-Learning Text}, Third Edition, Springer, New York.
	
	\bibitem{bookneuts}
	\textsc{Neuts, M. F.} (1994). \textit{Martix-geometric solutions in stochastic models. An algorithmic approach}, Mineola, NY: Dover Publications.
	
	
	
	\bibitem{Odd1995}
	\textsc{Odd Aalen, O.} (1995). \textit{Phase type distributions in survival analysis}, Scandinavian Journal of Statistics \textbf{22}(4): 447--463.
	
	\bibitem{Marshall2007}
	\textsc{Marshall, A.} and \textsc{Zenga, M.} (2009). \textit{ Simulating coxian phase-type distributions for patient survival}, International transactions in operational research \textbf{16}(2): 213--226.
	
	\bibitem{Nielsen2012}
	\textsc{Nielsen, B.} (2012). \textit{Lecture notes on phase-type distributions for 02407 Stochastic Process}.
	
	\bibitem{Asmussen1996}
	\textsc{Asmussen, S.}, \textsc{Nerman, O.} and \textsc{Olsson, M.} (1996). \textit{ Fitting phase-type distributions via the EM algorithm}, Scandinavian Journal of Statistics \textbf{23}(4): 419--441.
	
	\bibitem{Lin2007}
	\textsc{ Lin, X. Sh.} and \textsc{Liu, X.} (2007). \textit{Markov aging Process and Phase-Type Law of Mortality}, North American Actuarial Journal \textbf{11}(4): 92--109.
	
	\bibitem{Hougaard2000}
	\textsc{P Hougaard. Heidelberg} (2000). \textit{Analysis of Multivariate Survival Data}, Springer, 542 pages, ISBN: 0-387-98873-4.
	
	\bibitem{Dickdon2009}
	\textsc{David C. M. Dickson}, \textsc{Mary R. Hardy} and \textsc{Howard R. Waters} (2009). \textit{Actuarial mathematics for life contingent risks},Cambridge University Press, New York.
	
	\bibitem{Andersen1993}
	\textsc{Andersen, P.K.},\textsc{Borgan, O}, \textsc{Gill, R.D} and \textsc{ Keiding, N.} (1993). \textit{Statistical Models Based on Counting Processes}, Springer, New York.
	
	\bibitem{Meira2009}
	\textsc{Meira-Machado, L.}, \textsc{de Una-Alvarez, J.}, \textsc{Cadarso-Suarez, C.} and \textsc{Andersen, P.K.} (2009). \textit{Multi-state models for the analysis of time to event data}, Statistical Methods in Medical Research \textbf{18}: 195--222.
	
	\bibitem{Andersen2002}
	\textsc{Andersen PK} and \textsc{Keiding N.} (2002). \textit{Multi-state models for event history analysis}, Stat Methods Med Res \textbf{11}: 91--115.
	
	\bibitem{Andersen1982}
	\textsc{Andersen PK} and \textsc{Gill RD} (1982). \textit{Cox’s regression model for counting processes: a large sample study}, Ann Stat \textbf{10}: 1100–-20.
	
	
	
	
	\bibitem{Lin2000}
	\textsc{Lin DY}, \textsc{Wei LJ}, \textsc{Yang I} and \textsc{Ying Z} (2000). \textit{Semiparametric regression for the mean and rate functions of recurrent events}, J R Stat Soc B \textbf{62}:711–-30.
	
	\bibitem{Kelly2000}
	\textsc{Kelly PJ} and \textsc{ Lim L L-Y} (2000). \textit{Survival analysis for recurrent event data: an application to childhood infectious diseases}, Stat Med \textbf{19}: 13-–33.
	
	\bibitem{Cook2010}
	\textsc{Cook RJ} and \textsc{ Lawless JF} (2010). \textit{The Statistical analysis of recurrent events}, 2nd edn. New York, NY: Springer.
	
	\bibitem{Araujo2014}
	\textsc{Araujo A.}, \textsc{Meira-Machado L.} and \textsc{Roca-Pardinas J.} (2014).\textit{TPmsm: Estimation of the Transition Probabilities in 3-State Models.}, Journal of Statistical Software \textbf{62}(4): 1--29.
	
	\bibitem{Balboa2018}
	\textsc {Balboa V.} and \textsc{J. de Una-Alvarez} (2018). \textit{Estimation of Transition Probabilities for the Illness-Death Model: Package TP.idm}, Journal of Statistical Software \textbf{83}(10): 1--19.
	
	\bibitem{Efron1979}
	\textsc{ Efron, B.} (1979). \textit{Bootstrap methods: Another look at the Jackknife}, Ann. Statist \textbf{7}: 1--26.
	
	\bibitem{EfTib1993}
	\textsc{ Efron, B.} and \textsc{ Tibshiran, R.} (1993). \textit{An introduction to the bootstrap}, New York: Chapman \& Hall.
	\bibitem{HassanZadehetalldis}
	\textsc{Hassan Zadeh, A.} and \textsc{Jones, B.L.} and \textsc{Stanford, D. A.} (2014).\textit{	The use of phase-type models for disability insurance calculations}, Scandinavian Actuarial Journal. 2014(8): 714--728.
	
	

	
	
	\bibitem{Asghari2019}
	\textsc{ Asghari, R.} and \textsc{ Hassan zadeh, A.} (2020). \textit{Mortality modeling of skin cancer patients with actuarial applications}, Journal of North American Actuarial 0(0): 1--17, DOI:10.1080/10920277.2019.1670070.
	
	\bibitem{Kalb1980}
	\textsc{Kalbfleisch JD.} and \textsc{Prentice RL.} (1980). \textit{The Statistical Analysis of Failure Time Data}, John Wiley \& Sons.
	
	\bibitem{Crowley1977}
	\textsc{Crowley J.} and \textsc{Hu M.} (1977). \textit{ Covariance Analysis of Heart Transplant Survival Data}, Journal of the American Statistical Association \textbf{72}: 27--36.
	
	\bibitem{Faddy2000}
	\textsc{Faddy M.J.} and \textsc{McClean S.I.} (2000).\textit{ Analysing data on lengths of stay of hospital patients using phase-type distributions}, Applied Stochastic Models and Data Analysis.
	
	\bibitem{Ramaswami1999}
	\textsc{Latouche, G.}, \textsc{Ramaswami, V.} (1999). \textit{Introduction to Matrix Analytic Methods in Stochastic Modeling}, Series on statistics and
	applied probability. ASA-SIAM.
	
\end{thebibliography}
\end{document}